\documentclass[twocolumn,showpacs,preprintnumbers,amsmath,amssymb,superscriptaddress,nofootinbib,english]{revtex4}

\usepackage{epstopdf}
\usepackage{graphicx}
\usepackage{dcolumn}
\usepackage{bm}
\usepackage{epsfig}
\usepackage{graphicx}
\usepackage{hyperref}
\usepackage[usenames]{color}
\usepackage{url}
\usepackage{relsize}
\usepackage{comment}
\usepackage{tensor}

\hypersetup{
    colorlinks=true,
    linkcolor=green,
    citecolor=blue,
}

\newcommand{\remove}[1]{}

\def\be{\begin{equation}}
\def\ee{\end{equation}}
\def\ba{\begin{eqnarray}}
\def\ea{\end{eqnarray}}

\begin{document}

\title{Constraints on Disformal Couplings from the Properties of the Cosmic Microwave Background Radiation}
\author{Carsten van de Bruck}
\email[Email address: ]{C.vandeBruck@sheffield.ac.uk}
\affiliation{Consortium for Fundamental Physics, School of Mathematics and Statistics, University of Sheffield, Hounsfield Road, Sheffield, S3 7RH, United Kingdom}

\author{Jack Morrice}
\email[Email address: ]{app12jam@sheffield.ac.uk}
\affiliation{Consortium for Fundamental Physics, School of Mathematics and Statistics, University of Sheffield, Hounsfield Road, Sheffield, S3 7RH, United Kingdom}

\author{Susan Vu}
\email[Email address: ]{Susan.Vu@sheffield.ac.uk}
\affiliation{Consortium for Fundamental Physics, School of Mathematics and Statistics, University of Sheffield, Hounsfield Road, Sheffield, S3 7RH, United Kingdom}

\date{\today}

\begin{abstract}
Certain modified gravity theories predict the existence of an additional, non-conformally coupled scalar field. A disformal coupling of the field to the Cosmic Microwave Background (CMB) is shown to affect the evolution of the energy density in the radiation fluid and produces a modification of the distribution function of the CMB, which vanishes if photons and baryons couple in the same way to the scalar. We find the constraints on the couplings to matter and photons coming from the measurement of the CMB temperature evolution and from current upper limits on the $\mu$--distortion of the CMB spectrum. We also point out that the measured equation of state of photons differs from $w_\gamma = 1/3$ in the presence of disformal couplings. 
\end{abstract}

\maketitle

Among the extensions of General Relativity (GR) which have attracted a lot of attention in the literature are scalar--tensor theories, which are also well motivated by some extensions of  the standard model such as theories with extra spatial dimensions. In cosmology, those theories have been studied in the context of inflation and models for dark energy as they provide a natural mechanism to drive a period of accelerated expansion in the very early or late universe, see e.g. \cite{Copeland:2006wr, Amendola,Clifton:2011jh}. There are some open questions regarding scalar--tensor theories, such as whether the Einstein frame (EF) or Jordan frame (JF) formulation of a given theory is the most ``physical'', see e.g. \cite{Faraoni:1999hp} and references therein for a discussion. 

The EF is the frame in which the gravity and scalar sector look like GR plus a scalar field, but matter fields feel a metric which is usually conformally related to the EF metric $g_{\mu\nu}$. In the JF formulation of the theory, matter fields are uncoupled to the scalar, but the spin-2 and spin-0 fields are coupled in a non--trivial way. One could also consider more general transformations of the metric, in which the EF and the JF metrics are not related by a conformal transformation, but a so--called disformal transformation, which includes derivatives of the scalar field (see the definition in Eq. (\ref{gmunu}) below). Such possibilities were first conceived by Bekenstein \cite{Bekenstein:1992pj} and have attracted much attention since \cite{Kaloper:2003yf}-\cite{Zumalacarregui:2012us} and references therein. The disformal terms considered here are motivated for example in brane-world scenarios or theories of massive gravity.

In this paper, we consider scalar--tensor theories with disformally coupled scalar fields and study in particular the consequences on the properties of the CMB. We express observables in terms of the metric a (baryonic) observer feels, to side-step the question about a frame--independent formulation for disformally coupled fields. Frame--independence in conformally coupled theories has been discussed in \cite{Catena:2006bd,Chiba:2013mha}.

The theory we consider is given in the EF as 
\begin{eqnarray}\label{eframeaction}
{\cal S} &=& \int \sqrt{-g}d^4 x\left[ \frac{M_{\rm Pl}^2}{2}{\cal R}(g) -\frac{1}{2}g^{\mu\nu}(\partial_\mu\phi)(\partial_\nu\phi) - V(\phi) \right] \nonumber \\ 
&+& \sum S_{i} [{\tilde g}_{\mu\nu}^{i},\chi_i],
\end{eqnarray}
where the different matter fields $\chi_i$ propagate on geodesics for metrics given by
\be
\label{gmunu}
{\tilde g}_{\mu\nu}^{i} = C_{i}(\phi) g_{\mu\nu}+D_{i}(\phi)\partial_\mu \phi \partial_\nu \phi~.
\ee
From now on, reduced Planck units are assumed: $ c = \hbar = k_{B} = M_{\mathrm{Pl}} = 1 $.

We consider a cosmological setting for eq.~\eqref{eframeaction}, where $g_{\mu\nu}$ is the standard flat FRW metric, with $ds^2 = -dt^2 + a^2(t) \delta_{ij}dx^i dx^j$. Variation of the action with respect to the fields yields the Klein-Gordon equation for the scalar field $\phi$ and the energy conservation equations:
\begin{eqnarray}
\label{kg}
\ddot{\phi} + 3H\dot{\phi} + V' &=& \sum Q_{i}~,	\\
\label{rho_i} 
\dot{\rho}_{i} + 3H(\rho_{i} + p_{i}) &=& - Q_{i}\dot{\phi}~,
\end{eqnarray}
where the couplings are given by 
\begin{eqnarray}
\label{Q}
Q_{i} &=& \frac{C_{i}'}{2C_{i}} T_{i} + \frac{D_{i}'}{2C_{i}} \phi,_{\mu}\phi,_{\nu}T^{\mu \nu}_{i} \nonumber \\
	& \quad & - \nabla_{\mu} \bigg( \frac{D_{i}}{C_{i}} \phi_{\nu}T^{\mu \nu}_{i} \bigg)~.
\end{eqnarray}
Primes denote derivatives with respect to $\phi$, dots with respect to $t$ and $H=\dot a / a$ is the Hubble parameter. 
The stress-energy-momentum tensor, $T^{\mu\nu}_{i}$, with $ T \indices{^\mu_\nu} = \mathrm{diag}( -\rho, p, p, p) $, is assumed to be that of a perfect fluid. To simplify our analysis without loss of generality, we consider two coupled fluids, i.e. $i=1,2$. Then, for a FRW cosmology with two species coupled to the scalar field, \eqref{Q} becomes
\begin{eqnarray}
Q_{1} &=& \frac{\mathcal{A}_{2}}{\mathcal{A}_{1} \mathcal{A}_{2} - D_{1} D_{2} \rho_{1} \rho_{2}} \bigg( \mathcal{B}_{1} - \frac{ \mathcal{B}_{2} D_{1} \rho_{1} }{\mathcal{A}_{2}} \bigg)	\\
Q_{2} &=& \frac{\mathcal{A}_{1}}{\mathcal{A}_{1} \mathcal{A}_{2} - D_{1} D_{2} \rho_{1} \rho_{2}} \bigg( \mathcal{B}_{2} - \frac{ \mathcal{B}_{1} D_{2} \rho_{2} }{\mathcal{A}_{1}} \bigg)
\end{eqnarray}
where
\begin{eqnarray}
\mathcal{A}_{i} &=& C_{i} + D_{i} ( \rho_{i} - \dot{\phi}^2 )~,	\nonumber \\
\mathcal{B}_{i} &=& \bigg[ \frac{C_{i}'}{2} \bigg(-1 + 3 \frac{p_i}{\rho_i} \bigg) - \frac{D_{i}'}{2} \dot{\phi}^2 \nonumber \\
&\quad& + D_i \left\{ 3H\bigg( 1+ \frac{p_i}{\rho_i} \bigg) \dot{\phi} + V' + \frac{C_{i}'}{C_{i}} \dot{\phi}^2 \right\} \bigg] \rho_{i}~.	\nonumber
\end{eqnarray}
The following potential is considered
\be
\label{potential}
V=V_0e^{-\lambda\phi}~,
\ee
to be concrete and we choose $\lambda=1$. We assume that the scalar field plays the role of dark energy today. In what follows, we want to study disformal effects on its own and therefore set $C_{i}(\phi) = 1 $. We furthermore treat the couplings $D_i$ as a constant energy scale: $D_{i}(\phi) = M_{i}^{-4}$.

In order to study the properties of the CMB, we first want to find a kinetic description and relate that to a fluid description. This allows us to derive the equations (\ref{rho_i}) from first principles. To keep the discussion simple, we consider a single fluid with arbitrary equation of state but our arguments are easily generalised to a model with multiple species. We will return to the case of multiple couplings again later. The disformal coupling for the fluid to the scalar field will be denoted by $M$. 

In the JF the fluid is not coupled to the scalar field and we consider a perfect fluid of the form 
\be
\label{jordan}
\tilde{T}^{\mu \nu} = ( \tilde{\rho} + \tilde{p} ) \tilde{u}^{\mu} \tilde{u}^{\nu} + \tilde{p} \tilde{g}^{\mu \nu}~.
\ee
Throughout this paper, all tilde quantities are contracted with $ \tilde{g}_{\mu \nu} $, whereas quantities without tildes are contracted with the EF metric, $ g_{\mu \nu} $.

The fluid obeys the standard conservation equation 
\be
\label{jordanconservation}
\tilde{\nabla}_{\mu} \tilde{T}^{\mu \nu} = 0~,
\ee
where the covariant derivative is compatible with $ \tilde{g}_{\mu \nu} $ (i.e. $\tilde{\nabla}^\mu \tilde{g}_{\mu \nu}=0$).

Assuming that the scalar field is time--dependent only, as it is the case for a spatially homogeneous and isotropic space-time, one finds the relationship between the Einstein and Jordan frame 
energy momentum tensor \cite{Zumalacarregui:2012us}
\be
\label{mota}
T^{\mu \nu} = \sqrt{ 1 - \frac{\dot{\phi}^2}{M^4} } \tilde{T}^{\mu \nu}~,
\ee
where the derivative is taken with respect to $t$ with $g_{00} = -1$. This relation can be used to show that the equation of state in the different frames are related by 
\be\label{eos}
\frac{p}{\rho} = \frac{\tilde{p}}{\tilde{\rho}} \bigg(  1 - \frac{\dot{\phi}^2}{M^4}  \bigg)~.
\ee
From eqns.~\eqref{jordan}, \eqref{jordanconservation}, \eqref{mota} and \eqref{eos}, one can rederive eq.~\eqref{rho_i} for a single fluid. 

From a microscopic perspective, the well known definitions of the energy-momentum tensor of a fluid can be written as integrals over phase space. In the EF and JF we have respectively
\begin{eqnarray} 
T^{\mu\nu} = \int \frac{d^3 P}{\sqrt{-g}} \frac{P^\mu P^\nu}{P^0} f~, \label{em1}\\
{\tilde T}^{\mu\nu} = \int \frac{d^3 {\tilde P}}{\sqrt{-{\tilde g}}} \frac{{\tilde P}^\mu {\tilde P}^\nu}{\tilde P^0} {\tilde f}~, \label{em2}
\end{eqnarray}
where $P^\mu = dx^\mu/ d\lambda$ and ${\tilde P}^\mu = dx^\mu / d\tilde\lambda$ are the photon momenta, and $\lambda$ and $\tilde\lambda$ are affine parameters, in the two frames. We define the distribution functions $f$ and $\tilde f$ as usual by 
\begin{equation}
\label{neinstein}
d{\tilde N} = dx^1 dx^2 dx^3 d{\tilde P}_1 d{\tilde P}_2 d{\tilde P}_3 {\tilde f}
\end{equation}
and 
\begin{equation}
\label{njordan}
dN = dx^1 dx^2 dx^3 dP_1 dP_2 dP_3 f~.
\end{equation}

Combining eqns. \eqref{mota}, \eqref{em1}, \eqref{em2}, \eqref{neinstein} and \eqref{njordan}
with the condition that there are no collisions in the JF i.e. $ d\tilde{f}/dt = 0 $, we find the evolution of the EF distribution function, $f$, is
\be
\label{df/dt}
\frac{d}{dt}\bigg( \frac{d \tilde{\lambda}}{ d \lambda } f \bigg) = 0~.
\ee

Since we are interested in the properties of the CMB, we focus on radiation in the following. The EF Boltzmann equation can then be derived from eq.~\eqref{df/dt}, taking into account that photons now follow disformal geodesics 
\begin{equation}
\label{geodesic}
\tilde{P}^{\mu} \tilde{\nabla}_{\mu} \tilde{P}^{\nu} = 0^{\nu}~, \qquad 
			\tilde{g}_{\mu \nu} \tilde{P}^{\mu} \tilde{P}^{\nu} = 0~,	\\
\end{equation}
and that $ d \tilde{\lambda}/d \lambda $ does not necessarily vanish here. Such a result is already common in conformal theories. In fact we find $ d \tilde{\lambda}/d \lambda = ( 1 - \dot{\phi}^2 / M^4 )^{1/2} $.

In cosmology then, neglecting spatial gradients, the Boltzmann equation is
\begin{eqnarray}
\label{simplebolt}
{\hat{\cal L}}f &=&
P^0\frac{\partial f}{\partial t} - 
H\delta_{ij}P^iP^j\frac{M^4}{M^4 - \dot{\phi}^2} \frac{\partial f}{\partial P^0} \nonumber
\\ 
&=& \frac{\dot{\phi}\ddot{\phi}}{M^4 - \dot{\phi}^2}P^0f~.
\end{eqnarray}

Integrating eq.~\eqref{simplebolt} over momenta again reproduces eq.~\eqref{rho_i}. We can also derive an equation for the particle number density $n$, 
\begin{eqnarray}
\dot n + 3 H n = - \frac{\dot F}{2F} n,
\end{eqnarray}
with $F = (d\tilde \lambda/d\lambda)^2$. Note that photons travel along geodesics defined by the metric $\tilde{g}_{\mu\nu}$, but not in the EF metric $g_{\mu\nu}$. This leads to an additional term present in the EF Boltzmann equation, which can be interpreted as an effective collision term, as it involves no derivatives of $f$. According to \cite{Lima}, the equation for $n$ and $\rho_\gamma$ imply that the CMB does not obey the adiabaticity condition, so that the CMB is not an equilibrium blackbody. 

To summarize this part, we have presented a perfect fluid model in a disformal geometry that is consistent across both frames and found a consistent kinetic description. Note that the equation of state is not frame--independent in the presence of disformal couplings (see eq. \eqref{eos}). If $\tilde{p} = 0$, the pressure vanishes in the EF too, but for photons 
$\tilde{p} = \tilde{\rho}/3$ but $p \neq \rho/3$ in the EF. Note also that the distribution function $f$ as defined is not a frame--independent quantity for disformal transformations. 

We now return to the case of two fluids separately coupled to the scalar field. One fluid will be radiation (CMB), the other fluid will be matter (baryonic and dark). Note that there are now three distinguished frames: the EF, in which the action takes the form of eq. \eqref{eframeaction}. Then there is the frame in which the radiation fluid is uncoupled, but in general matter is coupled to the scalar, which we call the radiation frame (RF). The gravity--scalar part of the action will have a highly-nonstandard form. And finally, there is the frame in which matter is uncoupled, but radiation in general is coupled. Again, the gravity--scalar sector will have a highly--nonstandard form. We call this last frame the JF. Note that if matter and radiation are coupled with the same strength, the RF and the JF coincide. It is clear from the previous discussion that the distribution function and the equation of state depends on the frame used and observables are therefore best expressed in the JF. We will perform the calculations in the EF, but transform to JF quantities to compare to data. 

With this in mind, we will now discuss constraints on the disformal theory as defined by eq. \eqref{eframeaction}, with photons coupled to the scalar field with coupling $M_\gamma^4$ and matter (baryonic and dark) with coupling $M_{\mathrm{m}}^4$. We express the observables in terms of the JF variables, by which we mean we treat $g_{\mu\nu}^{(\mathrm{m})}$ as the fundamental metric as this is the metric which defines geodesics for an observer.

One of the consequences of the theory is that the measured (dimensionless) speed of light is given by 
\begin{equation}
c_{\rm obs}^2 = 1 - \left( \frac{1}{M_\gamma^4} - \frac{1}{M_{\rm m}^4}\right)\left( \frac{d\phi}{d\tau_{\rm m}} \right)^2~ ,
\end{equation}
which follows from $g_{\mu\nu}^{(\gamma)}dx^\mu dx^\nu  = 0$. $\tau_{\rm m}$ is the proper time measured by an observer. The expression for $c_{\rm obs}$ 
can also be written in terms of EF variables as 
\be
c_{\mathrm{obs}}^2 = \frac{ 1 - \frac{ \dot{\phi}^2 }{ M^4_{\gamma} } }{ 1 - \frac{ \dot{\phi}^2 }{ M^4_{\mathrm{m}} } }~.
\ee
We will need this expression below. Note that if $M_{\rm m} = M_\gamma$, the measured speed of light is equal to one in natural units. 

To constrain the theory, we use two observables: the temperature evolution of the CMB as a function of redshift and the spectral distortion generated by the disformal couplings. The temperature evolution of the CMB has been measured in the redshift range $0\leq z \leq 3$ by various authors \cite{Mather:1998gm,Avgoustidis:2011aa, Luzzi:2009ae, Noterdaeme:2010tm}. We associate an effective temperature to the JF energy density of the CMB , $ \rho_{\gamma}^{\mathrm{(m)}}$, in order to compare to these measurements. We relate this energy density to an effective temperature $T$ as $ \rho_{\gamma}^{\mathrm{(m)}} \propto T^4 $ \cite{Chluba:2011hw}. The CMB temperature evolution and the spectral distortion have been used in the past to constrain other theories beyond the standard model (see e.g. \cite{Lima} and \cite{Raffelt}). 

Eq.~\eqref{df/dt} states that the distribution function $f$ alone is not conserved, but that $f (d\tilde\lambda/d\lambda) = {\rm const}$. To be concrete, we assume that when the CMB is produced, the distribution function is of blackbody form to a very high accuracy with $\dot\phi = 0$, which is the case we consider here (therefore, we neglect any spectral distortions created in the early universe). The observed distribution function is given by $f_{\rm obs} = f_{\rm ini} (d\lambda/d\tilde\lambda)_0$ which we write in the form $f_{\rm obs} = (\exp(\nu/T + \mu) - 1)^{-1}$, where $\mu$ is an effective chemical potential \cite{jensnote}. Then, assuming that $\mu\ll 1$, we find from eq. \eqref{df/dt}
\be
\label{mudist}
\mu = \left( c_{\mathrm{obs}} - 1 \right) ( 1 - e^{-\nu/T } )~.
\ee

Eq. \eqref{mudist} closely agrees with the expression found by \cite{Brax:2013nsa}, though we take a very different approach. Importantly, our expression states that if $M_\gamma = M_{\rm m}$, no chemical potential is created. Current limits set by COBE/FIRAS \cite{Fixsen:1996nj} of $ |\mu| < 9 \times 10^{-5} $ will place constraints on the $M_{\mathrm{m}} \times M_{\gamma}$ parameter space. 

Finally, we need an expression for the redshift which is defined as \cite{Uzan}
\be
\label{redshift}
 1 + z = \frac{(u_{\mu} P^{\mu})_{\mathrm{em}}^{(\mathrm{m})} }{ (u_{\mu} P^{\mu})_{\mathrm{obs}}^{(\mathrm{m})} }~,
\ee
where the 4-velocity of the observer $ u_{\mu} = ( -c_{\mathrm{obs}}, 0 ,0 ,0) $, and $ P^{\mu} $ is the measured photon momentum. Both are JF quantities. 
Eq.~\eqref{geodesic} states that photons follow disformal geodesics; these equations can be rewritten in terms of the observer's frame by using the relation $ P^{\mu}_{(\gamma)} = P^{\mu}_{(\mathrm{m})} d\lambda_m / d\lambda_\gamma   =P^{\mu}_{(\mathrm{m})} c_{\mathrm{obs}}^{-1} $. This in combination with eq.~\eqref{redshift} leads to the redshift relation $ 1 + z = a_0/a $, as in GR.

The equations for matter, radiation (eq. \eqref{rho_i}) and the Klein--Gordon equation (eq. \eqref{kg}) are integrated to obtain a present day universe with density parameters $\Omega_{\rm m,0} = 0.3$ (cold dark matter and baryons), $\Omega_{\phi,0}=0.7$ (dark energy) \cite{Komatsu:2010fb} and $T_0=2.725$~K (CMB temperature) \cite{Mather:1998gm}. The field's initial value is fixed at $\phi_{\rm ini}=1.5~{\rm M}_{\rm Pl}$, while $V_0$ is allowed to vary.

\begin{figure}
\begin{center}
\scalebox{0.47}{\includegraphics{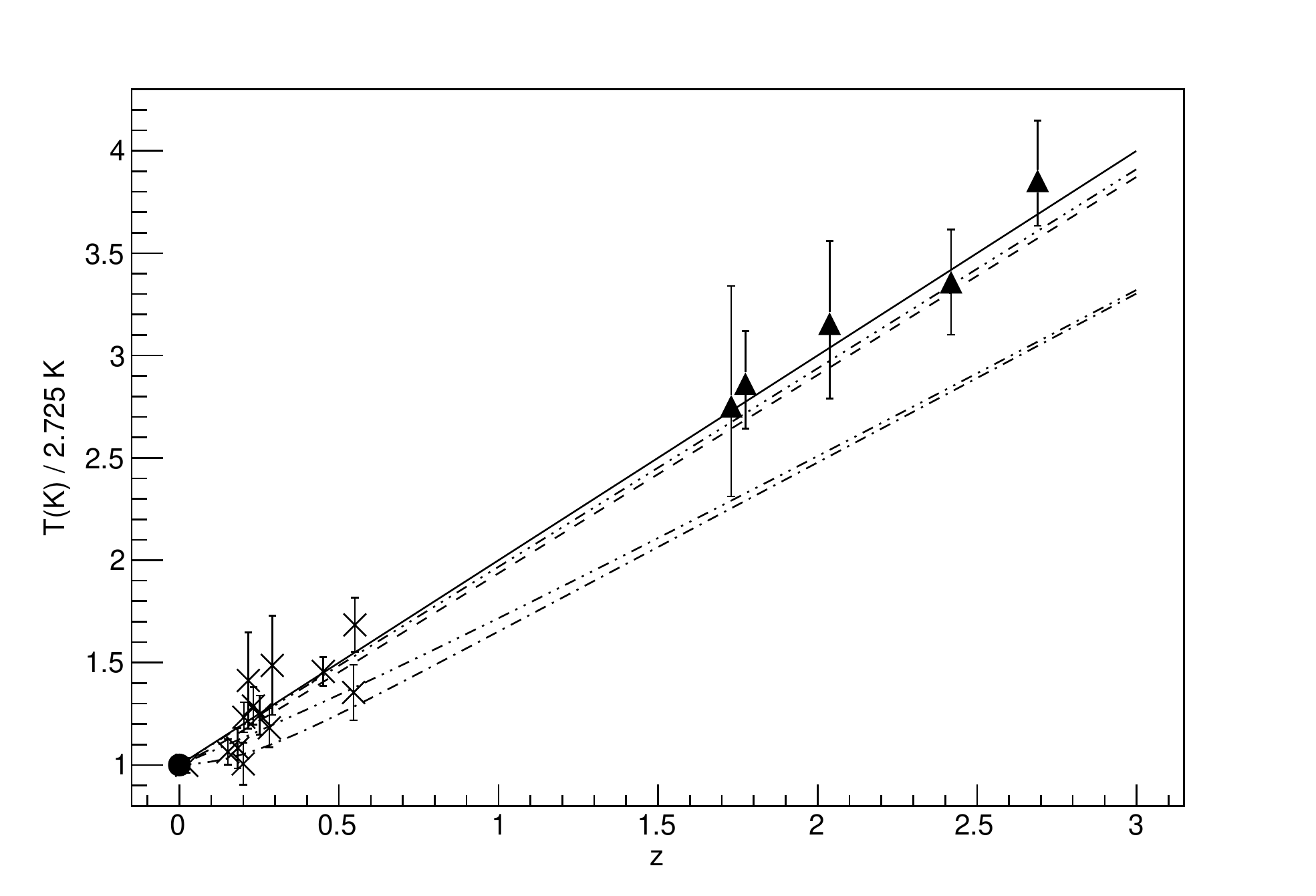}}
\caption{Plot of the ratio $T(z)/T_0$ where $T_0 = 2.725 \mathrm{K} $, against redshift, $ z $, for the exponential potential. The black circle marked at $ z = 0 $ is measured by COBE \cite{Mather:1998gm}. Different sets of measurements and their corresponding errors: Those marked with crosses are given by \cite{Luzzi:2009ae} and triangles are by \cite{Noterdaeme:2010tm}. 
There are five lines marked on this plot and each has an associated $ M_{\gamma} $: Solid line is $ M_{\gamma} \rightarrow 0, \infty $, dashed line $M_{\gamma}=2.203\times10^{-5}$~eV, dashed line with one dot $M_{\gamma}=3\times10^{-5}$~eV, dashed line with two dots $M_{\gamma}=1.5\times10^{-3}$~eV and dashed line with three dots $M_{\gamma}=2.2188\times10^{-3}$~eV. For this plot, we fix $M_{\rm m} = 0.05$~eV.}
\label{temperature}
\end{center}
\end{figure}

We have then calculated the expected evolution of $T(z)$ (equivalent to the evolution of $\rho_{\gamma}^{(\mathrm{m})}$) as a function of redshift for various values of $M_{\gamma}$. In Fig.~\ref{temperature}, we display the numerical results for acceptable models and those which are in conflict with the data, together with the measurements \cite{Avgoustidis:2011aa, Luzzi:2009ae, Noterdaeme:2010tm, Mather:1998gm}. The prediction of General Relativity ($M_{\gamma}\rightarrow\infty$), is also included.

For large $ M_{\gamma} $, we find that the weak coupling of $ \phi $ to radiation distorts the linear temperature-redshift relation of GR. However, this coupling has a damping effect on the field that increases as the coupling increases (i.e. smaller values for $M_\gamma$). For very large couplings, the damping is so severe that $ \dot{\phi} \rightarrow 0 $. As disformal couplings are derivative couplings, they will inevitably vanish in this limit. This is evident from Fig.~\ref{temperature}.

\begin{figure}
\begin{center}
\scalebox{0.47}{\includegraphics{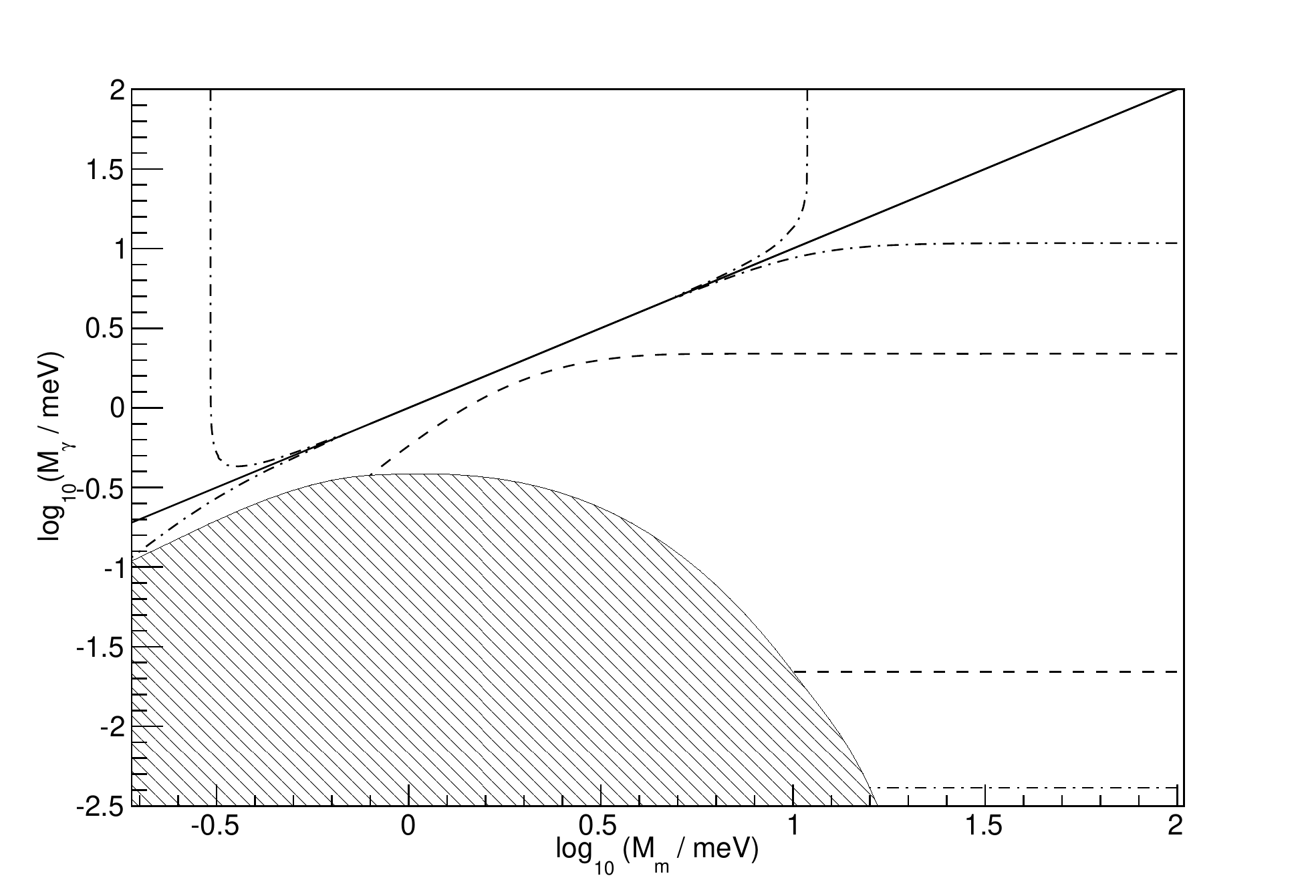}}
\caption{Bounds on the $M_{\mathrm{m}} \times M_{\gamma}$ parameter space. The solid line represents $M_{\mathrm{m}} = M_{\gamma}$. The region enclosed by the dashed lines is excluded by the CMB temperature evolution, the region shows the exclusion region above 68\% CL. The regions between the dashed--dotted lines are excluded by measured constraints on the $\mu$-distortions \cite{Fixsen:1996nj}. The shaded area is excluded from our search of the parameter space, as the numerics are unreliable in that region.}
\label{exclus}
\end{center}
\end{figure}

Through minimising chi-squared, an excluded region of the $M_{\mathrm{m}} \times M_{\gamma} $ parameter space is obtained. We find similar bounds by comparing the $\mu$-distortion in eq.~\eqref{mudist} with measured constraints from \cite{Fixsen:1996nj}, assuming frequencies of order $T_0$ \cite{Brax:2013nsa}. The bounds from both analyses are shown in Fig.~\ref{exclus}. We stress that though the constraints from $\mu$-distortions are stronger than temperature-redshift relation, both follow the same trend: a discrepancy between the couplings (i.e. $M_\gamma \neq M_{\rm m}$) is necessary for the disformal effects to be present. 

To conclude, we would like to highlight the following points: firstly, the equation of state for a given fluid is in general frame-dependent under disformal transformations. Secondly, the distribution function as defined in the standard way is frame--dependent too. Therefore, if in one frame the distribution function is of blackbody form, it is not so in another frame. As an application for the models considered here, we found a modification to the distribution function of the CMB photons, which vanishes if baryons and radiation couple with the same strength to the scalar field. We expressed observables in the JF, by which we mean the frame where baryons are uncoupled to the scalar field.  This is in contrast to \cite{Brax:2013nsa}, who work in the EF without referring to the JF.

Our work can be extended in several ways. To be completely general, $ M_{\mathrm{m}} $ and $ M_{\gamma}$ should be allowed to depend on the scalar field. In addition, it would be interesting to include conformal couplings to investigate how they affect the evolution of the scalar field and how the constraints on the parameters are then modified. As conformal terms require screening in higher density areas such as the solar system, disformal coupling scenarios with a screening mechanism must too be investigated, see e.g. \cite{Koivisto:2012za}. In such theories, $ \dot{\phi} $ is much smaller and we expect this to suppress disformal terms. Most importantly, a study of the evolution of perturbations is also warranted since the disformal coupling will also affect CMB anisotropies and growth of structure.

\acknowledgments The work of CvdB is supported by the Lancaster-Manchester-Sheffield Consortium for Fundamental Physics under STFC grant ST/J000418/1. SV is supported by an STFC doctoral fellowship. We thank Andrew Fowlie and Clare Burrage for useful discussions. We are grateful to Jens Chluba for important comments on possible spectral distortions of the blackbody spectrum in these theories.


\begin{thebibliography}{99}
\bibitem{jensnote} This effective chemical potential is created very late in the history of the universe ($z<10$) and not coming from scattering processes in the very early universe.
\end{thebibliography}
\end{document}